\documentclass[iop]{emulateapj}

\usepackage{braket}

\shorttitle{Impact Jetting, Magnetic Records of Chondrules, and Pebble Accretion}
\shortauthors{Hasegawa et al}

\begin{document}

\title{Forming Chondrites in a Solar Nebula with Magnetically Induced Turbulence}

\author{Yasuhiro Hasegawa\altaffilmark{1}, Neal J. Turner\altaffilmark{1}, Joseph Masiero\altaffilmark{1}, Shigeru Wakita\altaffilmark{2}, Yuji Matsumoto\altaffilmark{2}, Shoichi Oshino\altaffilmark{2}}
\affil{$^1$Jet Propulsion Laboratory, California Institute of Technology, Pasadena, CA 91109, USA}
\affil{$^2$Center for Computational Astrophysics, National Astronomical Observatory of Japan, Osawa, Mitaka, Tokyo 181-8588, Japan}
\email{yasuhiro@caltech.edu}

\begin{abstract}
Chondritic meteorites provide valuable opportunities to investigate the origins of the solar system.
We explore impact jetting as a mechanism of chondrule formation and 
subsequent pebble accretion as a mechanism of accreting chondrules onto parent bodies of chondrites,
and investigate how these two processes can account for the currently available meteoritic data.
We find that when the solar nebula is $\le 5$ times more massive than the minimum-mass solar nebula at $a \simeq 2-3$ AU
and parent bodies of chondrites are $\le 10^{24}$ g ($\le$ 500 km in radius) in the solar nebula,
impact jetting and subsequent pebble accretion can reproduce a number of properties of the meteoritic data.
The properties include the present asteroid belt mass, the formation timescale of chondrules, 
and the magnetic field strength of the nebula derived from chondrules in Semarkona.
Since this scenario requires a first generation of planetesimals that trigger impact jetting and serve as parent bodies to accrete chondrules,
the upper limit of parent bodies' masses leads to the following implications:
primordial asteroids that were originally $\ge 10^{24}$ g in mass were unlikely to contain chondrules,
while less massive primordial asteroids likely had a chondrule-rich surface layer.
The scenario developed from impact jetting and pebble accretion can therefore provide new insights into the origins of the solar system.
\end{abstract}


\section{Introduction} \label{intro}

Chondritic meteorites can be regarded as a fossil record of how the solar system formed \citep[e.g.,][]{dac14}.
The most invaluable records are contained in chondrules, which are the most abundant ingredient in chondrites.
Chondrules are millimeter-sized spherical particles 
that are formed as the consequence of transient heating events occurring in the solar nebula \citep[e.g.,][]{hcl05}.
Since such events may be relevant to important processes regulating the nebula evolution \citep[e.g.,][]{dmc12},
investigating chondrule formation can shed light on the origins of the solar system.
Chondrites themselves also possess profound information.
It is widely accepted that chondrites are fragments of planetesimals originating from the asteroid belt \citep[e.g.,][]{daw15}.
Taking into account the consideration that planetesimals are the building blocks of planets,
understanding of how chondrites are generated from their parent bodies can provide fundamental insight into 
the time evolution of the asteroid belt, planetesimal formation, 
and even (perhaps) the formation of (exo)planetary systems \citep[e.g.,][]{vzb14,jml15}.

Significant progress has recently been made on chondrule formation and accretion.
For instance, \citet{jmm15} have produced a pioneering work in which
planetesimal collisions and the resultant impact jetting are proposed as a mechanism of chondrule formation.
They have found that ejected materials from the collisional surface can reproduce the thermal history of chondrules,
and the collisional frequency during planetary accretion is high enough to account for the present asteroid belt mass \citep[also see][]{hwm15}.
In addition, an experimental study revealed that primitive chondrites and the chondrules embedded in them 
can retain the information of the magnetic field strength of the solar nebula \citep{fwl14}.
Furthermore,  \citet{lj12} have performed numerical simulations and 
shown that gas drag acting on pebbles can substantially accelerate the growth of massive bodies
due to the efficient accretion of pebbles onto the bodies \citep[also see][]{ok10,gio14,kl14}.
This process can be viewed as a mechanism of chondrule accretion onto parent bodies of chondrites \citep{jml15}.

Here we synthesize these recent steps on chondrule formation and accretion, 
and explore how impact jetting and subsequent pebble accretion can be consistent with the meteorite conditions.
We show below that the coupling of these processes with the meteorite data 
can provide tight constraints on the mass of the solar nebula at $a \simeq 2-3$ AU
and the size of parent bodies of chondrites in the solar nebula.
We will therefore conclude that both impact jetting and pebble accretion are intriguing processes 
for obtaining some valuable clues about the origins of the solar system.

\section{Chondrule formation and accretion} \label{mod}

\begin{table*}
\begin{minipage}{17cm}
\begin{center}
\caption{List of quantities}
\label{table1}
\begin{tabular}{clc}
\hline
Symbol           &   Meanings                                           & Value  \\ \hline 
$a$                 &  Semimajor axis       &   \\ 
$\Omega$       & Angular frequency   &  \\
$c_s$               &  Sound speed           &  \\
$\Sigma_{g}$  &   Gas surface density                    &            \\
$\Sigma_{s}$  &   Solid surface density                    &            \\
$\rho_{g}$       &   Gas volume density at the disk midplane &            \\
$f_d$               &   Increment factor                          &   \\ 
$T_d$              &   Disk temperature                        &            \\
$h_g$              &   Gas pressure scale height          &            \\
$\langle B \rangle $ & Magnetic field strength estimated from a chondrite   & 50-540 mG \\
$\langle B \rangle_{min} $ & Minimum value of $\langle B \rangle $  required for pebble accretion   &  \\
$\alpha_{\rm eff}$   & Effective value of $\alpha$ computed from $\langle B \rangle $  &  \\  
$\gamma_{turb}$  & Quantity related to the nature of turbulence     & 2  \\       \hline
$M_{p}$          &    Mass of protoplanets   &          \\
$R_p$            &    Radius of protoplanets &   \\
$\rho_p$        &    Mean density of protoplanets & 5 g cm$^{-3}$  \\
$M_{iso}$        &   Isolation mass of protoplanets   &          \\
$r_H$              &   Hill radius of protoplanets (= $ [ (2M_{p})/(3 M_{\odot}) ]^{1/3}a$ )  \\  \hline
$m_{pl}$         &   Mass of field planetesimals           & $10^{22}-10^{24}$ g     \\
$\rho_{pl}$        &  Mean density of field planetesimals & 2 g cm$^{-3}$  \\
$r_{pl}$           &  Radius of field planetesimals           & $\simeq 100-500$ km \\
$\langle e_{eq}^2 \rangle^{1/2} $ & Root mean square equilibrium eccentricity of field planetesimals in oligarchic growth       &   \\
$h_{pl}$         &   Vertical height of planetesimals  & \\ \hline
$r_{ch}$              & Characteristic size of chondrules                                                 &  1mm \\
$\rho_{ch}$        &  Bulk density of chondrules                                                           &  3.3 g cm$^{-3}$ \\
$h_{ch}$                & Scale height of chondrules                                                            &     \\
$\tau_{ch}$          &   Duration of the chondrule-forming epoch (since CAI formation)               &     $3-5 \times 10^6$ yr \\   
$m_{ch,IJ}$          &  Total mass of chondrules formed via impact jetting for $\tau_{ch}$ &   \\ 
$f_{IJ}$              &  Mass fraction of planetesimals that can eventually generate chondrules via impact jetting  & $10^{-2}$  \\
$\tau_{IJ}$       &   Timescale for forming chondrules via impact jetting    &  \\ 
\hline
\end{tabular}
\end{center}
\end{minipage}
\end{table*}

\subsection{Disk model}

We adopt the minimum-mass solar nebular (MMSN) model \citep{h81},
which can be given as (see Table \ref{table1} for the definition of variables),
\begin{equation}
 \Sigma_g = 1.7 \times 10^3 f_d  \left( \frac{a}{1 \mbox{ AU}} \right)^{-3/2} \mbox{ g cm}^{-2},
\end{equation}
\begin{equation}
 \Sigma_s = 7 f_d  \left( \frac{a}{1 \mbox{ AU}} \right)^{-3/2} \mbox{ g cm}^{-2},
\end{equation}
and
\begin{equation}
 T_{d} = 280  \left( \frac{a}{1 \mbox{ AU}} \right)^{-1/2} \mbox{K},
\end{equation}
where $f_d$ is a factor to examine the effect of disk mass on chondrule formation,\footnote{
Note that both the gas and the dust densities in \citet{hwm15} are increased by a factor of 1.5 for the standard model, following \citet{ki00}.}
and $T_d$ is derived under the optically thin limit.
Based on the above equations, we can obtain the following;
\begin{equation}
 \rho_g = \frac{\Sigma_g}{\sqrt{2 \pi} h_g} \simeq 1.4 \times 10^{-9} f_d  \left( \frac{a}{1 \mbox{ AU}} \right)^{-11/4} \mbox{ g cm}^{-3},
\end{equation}
and
\begin{equation}
 \frac{h_g}{a} = \frac{c_s}{a \Omega}\simeq 3.4 \times 10^{-2}   \left( \frac{a}{1 \mbox{ AU}} \right)^{1/4},
\end{equation}
where the vertical hydrostatic equilibrium is assumed 
and the mean molecular weight is taken as 2.3 in units of atomic hydrogen.
We adopt a stellar mass of $M_* = 1 M_{\odot}$.

\subsection{Impact jetting} \label{mod_ij}

We consider impact jetting as one of the primary processes for generating chondrules 
which can occur during planetary accretion.
For this case, the resultant abundance of chondrules can be characterized well by $f_{IJ}M_{iso}$
as long as formed protoplanets can reach $M_{iso}$ within a given disk lifetime \citep{hwm15}.\footnote{
The adopted value of $f_{IJ}$ may be a lower limit
because it is derived only from vertical impacts, with a conservative estimate of the threshold impact velocity \citep{jmm15}.}
This formula is obtained because chondrule-forming collisions can be realized 
when protoplanets undergo oligarchic growth and acquire most of their masses \citep[e.g.,][]{ki98}.

\begin{figure}
\begin{center}
\includegraphics[width=8cm]{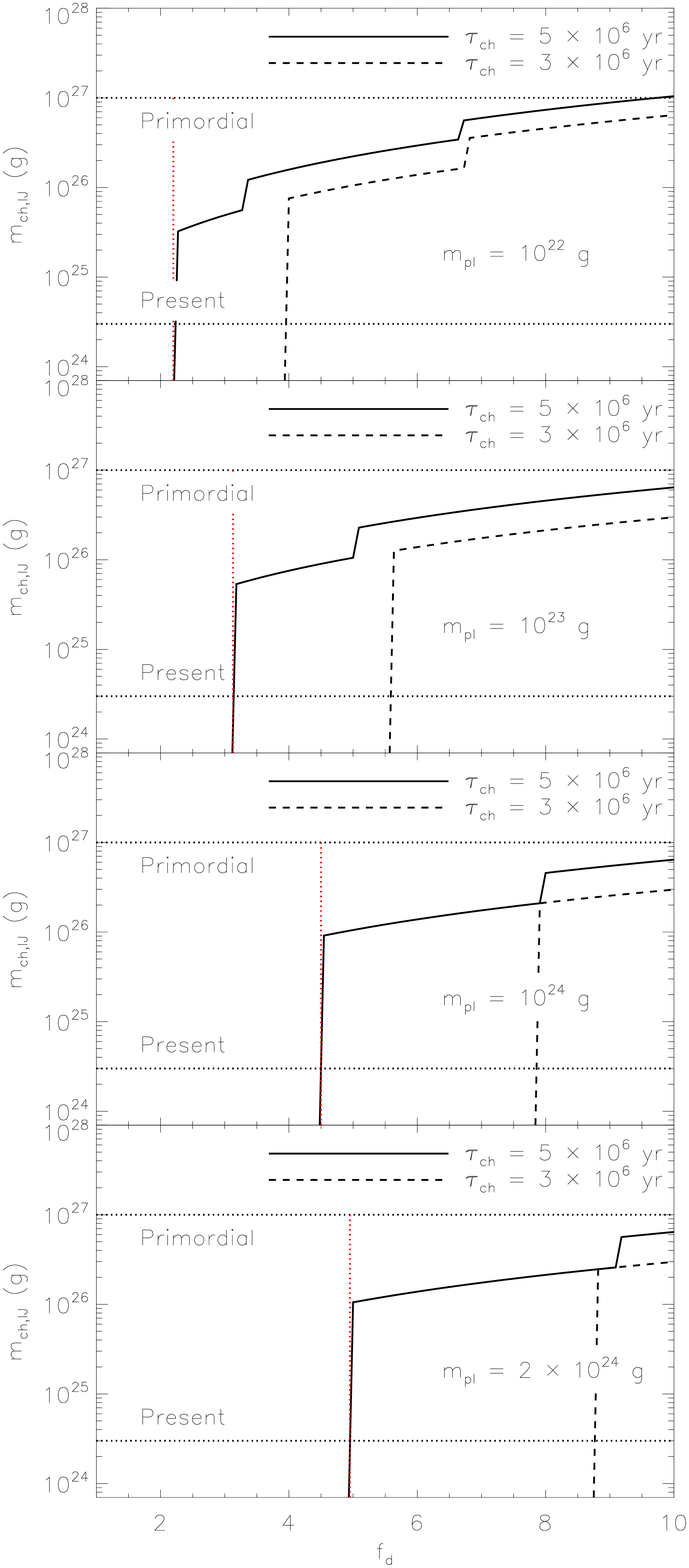}
\caption{Mass of chondrules ($m_{ch,IJ}$) produced by impact jetting via planetary accretion at $a=2-3$ AU 
as a function of $f_d$ (see Table \ref{table1}).
The value of $m_{pl}$ varies from $10^{22}$ g (top) to $2 \times 10^{24}$ g (bottom).
For comparison purposes, both the primordial and the present asteroid belt masses are plotted as the horizontal dotted lines.
The minimum value of $f_d$ above which $m_{ch,IJ}$ is large enough to reproduce the required abundance of chondrules
increases with increasing $m_{pl}$ (see the vertical dotted line).}
\label{fig1}
\end{center}
\end{figure}

The total abundance of chondrules for $2 \mbox{AU} \la a \la 3 \mbox{AU}$
can be readily computed as 
(see \citet{hwm15} for a complete discussion and references)
\begin{equation}
 \label{eq:m_ij}
 m_{ch,IJ}(\tau_{ch}) \simeq \sum_a f_{IJ} M_{iso} (a),
\end{equation}
where 
\begin{eqnarray}
 \label{eq:mp_iso}
 M_{iso}  & \simeq &  9.4 \times 10^2 M_{\oplus} 
                                     \left( \frac{\Sigma_s}{7 \mbox{ g cm}^{-2}} \right)^{3/2}  
                                     \left( \frac{a}{1 \mbox{ AU}} \right)^{3}.
\end{eqnarray}
Equation (\ref{eq:m_ij}) assumes that protoplanets are separated by their feeding zones ($\simeq 10 r_H$) with each other,
and that chondrules were formed over a time interval $\tau_{ch}$.
Since meteorite records indicate that 
chondrule formation began around the time of CAI formation and continued for $3-5 \times 10^6$ years \citep[e.g.,][]{cbk12,bcb15},
two characteristic values of $\tau_{ch}$ are considered ($\tau_{ch}=3 \times 10^6$ years and $\tau_{ch}=5 \times 10^6$ years, see Figure \ref{fig1}).
In the summation, terms contribute only
if $\tau_{ch}$ is longer than the time ($\tau_{IJ}$) for protoplanets  to grow up to $M_{iso}$,
which is given as \citep{im93}
\begin{equation}
\label{eq:tau_ij}
\tau_{IJ} \equiv  \left. f_{\tau} \frac{M_{p}}{ dM_{p} /dt } \right|  _{M_p = M_{iso}},
\end{equation}
where 
\begin{equation}
 \label{eq:dmdt}
 \frac{dM_{p}}{dt} \simeq 4 \pi \Sigma_s \frac{ G M_{p}R_p}{\langle e_{eq}^2 \rangle a^2 \Omega},
\end{equation}
and
\begin{eqnarray}
\label{eq:e_oli}
&\langle e_{eq}^2 \rangle^{1/2} & \simeq   2.8 \times 10^{-2} \left( \frac{m_{pl}}{10^{23} \mbox{ g}} \right)^{1/15}
                                               \left( \frac{\rho_{pl}}{2 \mbox{ g cm}^{-3}} \right)^{2/15}  \\ \nonumber
                            & \times  & \left( \frac{\rho_{g}}{1.4 \times10^{-9} \mbox{ g cm}^{-3}} \right)^{-1/5} 
                                               \left( \frac{a}{1 \mbox{ AU}} \right)^{-1/5}  \left( \frac{M_{p}}{0.1 M_{\oplus}} \right)^{1/3}.                 
\end{eqnarray}
As found by \citet{hwm15}, 
$f_{\tau} =3$ is needed to reproduce the results of more detailed calculations.\footnote{
In \citet{hwm15}, $\tau_{IJ}$ is labeled as $\tau_{end}$.}
Also, the oligarchic phase is assumed in computing planetesimals' eccentricity ($\langle e_{eq}^2 \rangle$),
wherein the balance between the pump up of $\langle e_{eq}^2 \rangle$ by a protoplanet and its damping by the disk gas is considered.
Thus, $m_{ch,IJ}$ is calculated based on the semi-analytical formulation developed in \citet{hwm15},
under the assumption that protoplanet-planetesimal collisions play a dominant role for $m_{ch,IJ}$. 

Figure \ref{fig1} shows the resultant value of $m_{ch,IJ}$ as a function of $f_d$.
The value of $m_{pl}$ varies from $10^{22}$ g to $2 \times 10^{24}$ g in each panel,
in order to examine its dependence on $m_{ch,IJ}$ (see equation (\ref{eq:e_oli})).
The results show that $m_{ch,IJ}$ is very low for small values of $f_d$,
and it rises suddenly as $f_d$ increases (see the vertical dotted line).
This occurs because a disk becomes massive enough at that time,
so that a growing protoplanet at $a = 2$ AU can reach $M_{iso}$ within given values of $\tau_{ch}$. 
Once $f_d$ exceeds these critical values, the increment of  $m_{ch,IJ}$ becomes gradual 
until other sudden jumps in $m_{ch,IJ}$ appear.
These jumps originate from an additional contribution arising from another protoplanet
that can now reach $M_{iso}$ within $\tau_{ch}$ by increasing $f_d$.
Note that this secondly formed protoplanet is separated from the first formed protoplanet by $\Delta a \simeq 0.1$ AU.

The results also show that the first sudden rises of $m_{ch,IJ}$ require a higher value of $f_d$
as $m_{pl}$ increases (see the vertical dotted line on each panel in Figure \ref{fig1}).
This is a simple reflection of $\tau_{IJ}$ that is an increasing function of $m_{pl}$ 
(see equations (\ref{eq:tau_ij}), (\ref{eq:dmdt}), and (\ref{eq:e_oli}));
when massive planetesimals are involved in planetary accretion,
the value of $\langle e_{eq}^2 \rangle$ becomes higher,
since the eccentricity damping by the disk gas becomes less effective.
Then, the growth rate of a protoplanet slows down.
As a result, a higher value of $f_d$ is needed to accelerate the formation of protoplanets, 
in order to satisfy the condition that protoplanets obtain $M_{iso}$ within $\tau_{ch}$.

Thus, the abundance of chondrules formed by impact jetting through planetesimal collisions 
is very likely significant enough to reproduce the present asteroid belt mass; 
for given values of $\tau_{ch}$,
the formation of protoplanets at $a \simeq 2$ AU generates a large amount of chondrules for reasonable values of $f_d$ and $m_{pl}$.
This finding is consistent with the previous studies \citep{jmm15,hwm15}.
Nonetheless, it is obvious that the results degenerate for certain ranges of $f_d$ and $m_{pl}$. 
We will determine the ranges in these two quantities below, 
using further constraints from meteorite measurements.

\subsection{Magnetic records of chondrules} \label{mod_alpha}

Magnetic fields can play an important role in evolution of circumstellar disks \citep[e.g.,][]{nfg14}.
The existence of these fields in the solar nebula is supported by
measurements of the Semarkona ordinary chondrite \citep{fwl14}.
In Fu et al's experiment, the thermoremanent magnetization of chondrules in the chondrite and 
the direction of magnetic fields are measured simultaneously.
These measurements suggest that 
some chondrules in this sample very likely recorded the magnetic field strength
that they experienced well before being accreted  by their parent body.
In other words, the data trace the magnetic field strength of the solar nebula around the chondrule-forming region,
which is $\langle B \rangle  \simeq 50-540$ mG.\footnote{The estimate is based on measurements from eight chondrules,
each of which might have recorded the magnetic field strength at different locations and times.}
This information is of fundamental importance for characterizing the strength of turbulence in the nebula,
which can be computed as \citep[e.g.,][]{w07,oh11}
\begin{equation}
\label{eq:alpha_eff}
\alpha_{\rm eff} = \frac{ \langle B_r B_{\phi} \rangle }{ \Sigma_g h_g \Omega^2 } \le \frac{ \langle B \rangle ^2}{ \Sigma_g h_g \Omega^2 },
\end{equation}
where $B_r$ and $B_{\phi}$ are the radial and the azimuthal components of magnetic fields, respectively, 
and the famous $\alpha$-prescription is adopted to label the turbulent strength \citep{ss73}.
We will use the resultant value of $\alpha_{\rm eff}$ (and $\langle B \rangle$) below.

\subsection{Pebble accretion} \label{mod_peb}

Pebble accretion is a promising candidate process for forming the chondrite parent bodies \citep[e.g.,][]{lj12,jml15}.
For this case, 
the vertical scale heights of chondrules ($h_{ch}$) and the planetesimals that accrete them ($h_{pl}$)
are important for determining the accretion efficiency.
Specifically, pebble accretion is most efficient for planetesimals with a scale height $h_{pl} < h_{ch}$,
namely, when they lie among the pebbles \citep{lkd15}.
We assume this holds in the following.
Note that $\tau_{ch}$ ($\sim 3-5 \times10^6$ years) is longer than 
the timescale at which a steady state of $h_{pl}$ is achieved,
which is less than $10^6$ years \citep{ki00}.

Chondrules are typically about 1 mm in size,
and large enough to settle toward the midplane due to stellar gravity.
In addition, the solar nebula is very likely turbulent,
which diffuses chondrules vertically.
Based on equation (\ref{eq:alpha_eff}), 
$\alpha_{\rm eff} \simeq 10^{-4}-10^{-2}$ for $f_d \ga 4$ at $ a \simeq 2-3$ AU.
This is a manifestation that shows that magnetic field can play an important role in the solar nebula.
As a result, $h_{ch}$ can be determined by the balance between stirring and settling.
Then, $h_{ch}$ can be written as \citep{dms95}
 \begin{equation}
 \frac{h_{ch}(r_{ch})}{h_g}= \frac{ H} {\sqrt{ 1+ H ^2}},
\label{eq:h_d}
\end{equation}  
where 
\begin{equation}
 \label{eq:bar_h}
 H  = \left(\frac{1}{1+\gamma_{turb}} \right)^{1/4} \sqrt{\frac{\alpha_{\rm eff} \Sigma_g}{\sqrt{2\pi} \rho_{ch} r_{ch}}}.
\end{equation}
This formula is consistent with other studies \citep[e.g.,][]{yl07},
and MHD simulations show that this approach works well for turbulent regions \citep[e.g.,][]{fp06}.

For the planetesimals ($h_{pl}$),
an eccentricity pump-up by a protoplanet is very likely to dominate 
over the effect of the random torque caused by turbulence in the solar nebula.
This is because $\langle e_{eq}^2 \rangle^{1/2} \sim 10^{-1}-10^{-2}$ for the former, 
while $\langle e_{eq}^2 \rangle^{1/2} \sim 10^{-2}-10^{-3}$ for the latter,
given that $\alpha_{\rm eff} \simeq 10^{-2}-10^{-4}$ \citep[e.g.,][]{igm08,gnt12}.
If this is the case, 
\begin{equation}
h_{pl} \simeq a \langle e_{eq}^2 \rangle^{1/2}  / 2,
\end{equation}
under the assumption that the planetesimals' random motions reach equipartition between eccentricity and inclination.
Note that $h_{pl}$ depends both on $f_d$ and on $m_{pl}$ (see equation (\ref{eq:e_oli})).
Then, we can find out the minimum value of magnetic fields ($\langle B \rangle_{min}$) 
above which pebbles are stirred enough to make their accretion efficient
by equating $h_{ch}$ with $h_{pl}$ (using equation (\ref{eq:alpha_eff}));
\begin{eqnarray}
\label{eq:b_min}
\langle B \rangle & \ge &  \left( 2 \pi (1+\gamma_{turb}) \right)^{1/4}  
                                         \sqrt{\rho_{ch} r_{ch} h_g \Omega^2 \frac{ (h_{pl}/h_g)^2 }{ 1 - (h_{pl}/h_g)^2 } } \\ \nonumber
                           &  \equiv &  \langle B \rangle_{min}.
\end{eqnarray}
Thus, we can compare $\langle B \rangle_{min}$ with that derived from the chondrite (see Section \ref{mod_const}).

\begin{figure}
\begin{center}
\includegraphics[width=8cm]{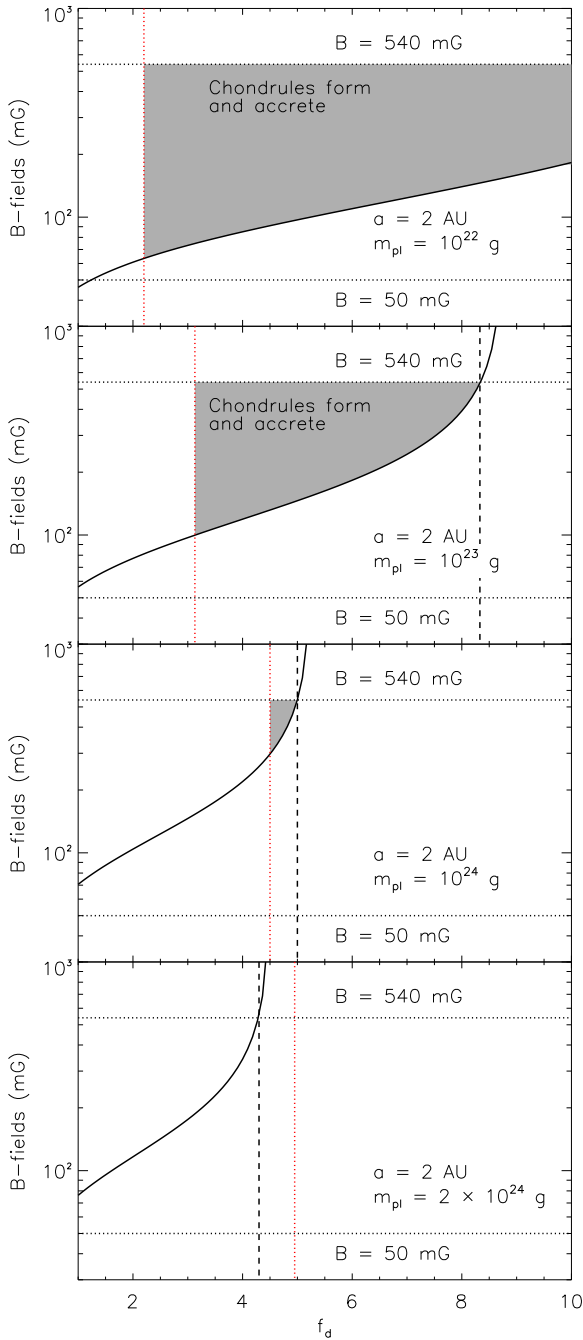}
\caption{Minimum value of the magnetic field strength ($\langle B \rangle_{min}$) threading the solar nebula,
which is needed to achieve efficient chondrule accretion onto planetesimals, 
as a function of $f_d$ (see Table \ref{table1}).
The value of $m_{pl}$ varies on each panel (as Figure \ref{fig1}).
The vertical dashed line denotes a critical value of $f_d$
above which chondrule accretion onto planetesimals is not realized.
Coupling with the constraint derived from the chondrule abundance 
(see the vertical dotted line, which is identical to the one in Figure \ref{fig1}),
the results show that all the requirements inferred from the currently available meteoritic samples can be met
when $f_d \le 5$ and $m_{pl} \le 10^{24}$ g  at $a =2$ AU (see the hatched region).}
\label{fig2}
\end{center}
\end{figure}

Figure \ref{fig2} shows the results of  $\langle B \rangle_{min}$ as a function of both $f_d$ and $m_{pl}$.
The calculations are done at $a=2$ AU, because this is where the main contribution to form chondrules occurs (see Section \ref{mod_ij}).
We find that $\langle B \rangle_{min}$ increases with increasing $f_d$.
This trend can be understood as the following.
As a disk becomes more massive (a higher value of $f_d$),
the value of $M_{iso}$ also becomes higher.
This in turn pumps up the planetesimals' eccentricities ($\langle e_{eq}^2 \rangle$) and their scale heights ($h_{pl}$).
Consequently, a higher value of $\langle B \rangle_{min}$ is needed to fulfill the condition that $h_{ch} = h_{pl}$.
For the dependence of $m_{pl}$ on $\langle B \rangle_{min}$ (see each panel),
the results show that when massive planetesimals are considered,
the required value of $\langle B \rangle_{min}$ becomes higher for a given value of $f_d$.
This is simply because $h_{pl}$ is an increasing function of $m_{pl}$;
the eccentricity damping by the disk gas becomes weaker for massive planetesimals (see equation (\ref{eq:e_oli})).
As a result, $\langle B \rangle_{min}$ should be stronger in order to expand the pebble sea for the vertical direction
to catch up with the increment of $h_{pl}$ by increasing $m_{pl}$.

Thus, the coupling of pebble accretion with the dynamics of planetesimals that is regulated by growing protoplanets 
enables an estimate of magnetic fields that are needed to achieve efficient pebble accretion in the system.

\subsection{New constraints on $f_d$ and $m_{pl}$} \label{mod_const}

We are now in a position to derive new constraints on the solar nebula and parent bodies of chondrites,
under the assumption that impact jetting and subsequent pebble accretion play a dominant role in forming the bodies.

Based on the discussion in Section \ref{mod_ij},
a lower limit of $f_d$ is obtained. 
Above this limit both the abundance of chondrules and their formation timescales are consistent with the meteoritic data
(see the vertical dotted line in Figure \ref{fig1}).
An upper limit of $f_{d}$ comes from comparing  $\langle B \rangle_{min}$ with the magnetic fields inferred from Semarkona 
(see the vertical dashed line in Figure \ref{fig2}, also see Section \ref{mod_peb}).
In the end, the most likely values of $f_d$ can be specified as a function of $m_{pl}$ 
(see the hatched region in Figure \ref{fig2}).
It is important that the hatched region shrinks as $m_{pl}$ increases,
and it eventually disappears for the case in which $m_{pl} > 10^{24}$ g.

Our results therefore demonstrate that 
the combination of impact jetting with pebble accretion can satisfy all the currently available meteoritic data
when the solar nebula is $\le 5$ times more massive than the MMSN at $a \simeq 2-3$ AU 
and parent bodies of chondrite are $\le 10^{24}$ g in the solar nebula.

\section{Discussion} \label{disc}
 
So far we have assumed that chondrules generated from forming protoplanets via impact jetting 
are accreted by the surrounding field planetesimals as long as $h_{pl} < h_{ch}$.
What happens if the condition that $h_{pl} < h_{ch}$ is not satisfied?
For this case, it is expected that the majority of chondrules will be accreted by the protoplanets and/or the central star.
In the current asteroid belt, chondrules are apparently ubiquitous.
In addition, protoplanets are too large to be disrupted to form asteroids.
Furthermore, it is obvious that smaller planetesimals collectively have a larger surface area than more massive bodies \citep{jmm15}.
Thus, it is necessary that (at least) some fractions of chondrules should be accreted by small planetesimals.

Our results can be compatible with several previous findings.
\citet{mbn09} suggest that 
the planetesimals originally at the asteroid belt's location were at least 50 km in radius.
This planetesimal radius or even a higher value (up to 500 km in radius) is needed 
to reproduce a bump observed at the planetesimal radius of $\sim 50$ km in the size distribution of the current asteroid belt.
In addition, \citet{jml15} have recently shown that 
planetesimals with a characteristic radius of $\sim 100$ km can experience efficient pebble accretion within $\sim 3 \times 10^{6}$ years.
Since planetesimals that are smaller than $50-100$ km in radius can be generated as fragments of larger bodies \citep[e.g.,][]{mbn09,daw15},
their results may provide a lower limit of parent bodies' mass.
If this is the case, 
coupling of our results with theirs enables the identification of a certain range of $m_{pl}$ for primordial asteroids.
Furthermore, a higher value of $f_d (> 1)$ is consistent with 
the finding derived from the standard core accretion picture applied to the solar system \citep[e.g.,][]{p96,hbl05}
as well as the populations of exoplanets \citep[e.g.,][]{tmr08,hp13a}; 
more massive disks than the MMSN are required to form Jupiter and Saturn.

Another important implication derived from our results may be the mass segregation of parent bodies in terms of chondrule inclusion.
In our scenario, the presence of a first generation of planetesimals is needed both for impact jetting and for pebble accretion.
Then, the resultant constraint on parent bodies' size may imply that
chondrules might have had less of a chance to be accreted by primordial asteroids that were $\ge 500$ km in radius,
while smaller primordial asteroids likely had a surface layer in which chondrules were abundant.
It is also interesting that the new constraint on $m_{pl} (\le 10^{24}$ g) is roughly comparable to the mass of Vesta/Ceres.
A careful examination of Vesta/Ceres may lead to acquisition of further clues about our scenario.

It is clear that a more detailed investigation is needed to verify the picture discussed here.
For instance, our scenario requires a first generation of planetesimals as discussed above.
Possible mechanisms to form such planetesimals \citep[i.e., streaming instabilities,][]{jomk07} should be further explored in this context.
Also, detailed studies of how the resultant chondrules can be accreted onto planetesimals 
under the presence of a protoplanet should be undertaken.
One of the key steps may be to perform numerical simulations that treat the random torque on protoplanets from turbulent disks,
which may affect how and where chondrules form and accrete.
Finally, our current setup is local, that is, it is isolated entirely from its surrounding.
It is important to examine how the picture developed here can be affected by the presence of nearby planets such as Jupiter.
These kind of global simulations may show how our scenario could favor or disfavor the Grand-tack scenario \citep{wmr11},
which proposes that a mixture of volatile-rich and volatile-poor asteroids in the current asteroid belt 
may have originated from gas-induced migration of Jupiter and Saturn.

In the next few years, potential sample-return missions and laboratory studies should yield 
a broader and deeper view of the history recorded in primitive asteroids.
This will enable tests of the ideas that chondrules formed through impact jetting, cooled in the asteroid belt's ambient magnetic field, 
and were quickly swept up by planetesimals with the aid of gas drag forces.


\acknowledgments

The authors thank Mario Flock, Katherine Kretke, Ryuji Morishima, and Satoshi Okuzumi for stimulating discussions,
and an anonymous referee for useful comments on our manuscript.
This research was carried out at JPL/Caltech, under a contract with NASA.
Y.H. is supported by JPL/Caltech.






\bibliographystyle{apj}

\end{document}